\input{aipcheck}

\documentclass[
    ,final            
  ]
  {aipproc}

\layoutstyle{6x9}

\begin{document}

\title{NC $\pi^0$ Production in the MiniBooNE Antineutrino Data}


\author{V. T. Nguyen, for the MiniBooNE Collaboration} {
  address={Department of Physics, Columbia University}
}

\vspace{-0.2cm}
\begin{abstract}
The single largest background to future $\bar{\nu_{\mu}}\rightarrow \bar{\nu_e}$ ($\nu_{\mu} \rightarrow {\nu_e}$) oscillation searches is neutral current $\pi^{0}$ production.  MiniBooNE, which began taking antineutrino data in January 2006, has the world's largest sample of $\pi^{0}$'s produced by antineutrinos in the 1 GeV energy range.  These neutral pions are primarily produced through the $\Delta$ resonance but can also be created through ``coherent production.''  The latter process is the coherent sum of glancing scatters of (anti)neutrinos off a neutron or proton, in which the nucleus is kept intact but a $\pi^{0}$ is created.  Current analysis of NC $\pi^0$ production in the MiniBooNE antineutrino data will be discussed.
\end{abstract}

\maketitle


\vspace{-1.cm}
\section{The MiniBooNE Experiment}
\vspace{-.1cm}

The Mini Booster Neutrino Experiment (MiniBooNE) [1] uses the Fermilab Booster neutrino beam, which is produced from 8 GeV protons incident on a Be target.  The target is located inside a horn, which produces a toroidal magnetic field that focuses pions and kaons.  These mesons then pass through a collimator and can decay inside a 50-m-long tunnel.   The MiniBooNE Cherenkov detector, located 541 m from the front of the target, is a spherical tank filled with 800 tons of pure mineral oil. The inside wall is lined with 1280 phototubes while the outside wall is covered by 240 veto phototubes.

MiniBooNE began running in antineutrino mode in January 2006 and currently has the world's largest sample of $\pi^0$'s produced by antineutrinos in the 1 GeV energy range (with over 1700 events after cuts).  As with any antineutrino beam, there is a non-negligible ``wrong-sign'' (WS) $\nu$ background, which comprises $\sim30\%$ of the total events.  This is due to the large $\pi^+$ to $\pi^-$ ratio produced by a proton beam incident on the target.

\vspace{-0.4cm}
\section{Neutral Current $\pi^0$ Production}

\vspace{-.1cm}
At low energy, neutral current (NC) $\pi^0$'s are produced via two different mechanisms:
\begin{equation}
\bar{\nu}N \rightarrow \bar{\nu} \Delta \rightarrow \bar{\nu}\pi^0 N  \qquad (resonant)
\end{equation}
\vspace{-.7cm}
\begin{equation}
\bar{\nu}A \rightarrow \bar{\nu} A \pi^0  \qquad (coherent)
\end{equation}

In resonant $\pi^0$ production, a (anti)neutrino interacts with an n or a p, exciting the nucleon into a $\Delta^0$ or $\Delta^+$, which then decays back to an n or a p, emitting a $\pi^0$.  In coherent $\pi^0$ production, very little energy is exchanged between the (anti)neutrino and the n or p.  The nucleus is left intact but a $\pi^0$ is created from the coherent sum of scattering from all the nucleons.  A signature of this process is a $\pi^0$ which is distinctly forward-scattered.

\vspace{-0.1cm}
\subsection{NC $\pi^0$'s as an Oscillation Search Background}
A $\pi^0$ decays very promptly into two photons ($\tau_{\pi^0}\sim8$x$10^-17$s), which can be detected in MiniBooNE as two electron-like rings.  However, if only one track is resolvable, this event can be misidentified as a single electron-like ring, which is the event signature of a $\bar{\nu_e}$ ($\nu_e$) interaction.  Thus, understanding of NC $\pi^0$ events is crucial.

Unfortunately, there is currently only one published measurement of the absolute rate of antineutrino NC $\pi^0$ production [2]; this measurement was reported with 25$\%$ uncertainty at 2 GeV.  Furthermore, current theoretical models on coherent $\pi^0$ production can vary by up to an order of magnitude in their predictions at low energy, the region most relevant for (anti)neutrino oscillation experiments.

\vspace{-0.1cm}
\section{Coherent $\pi^0$'s in Terms of $E_\pi(1-\cos{\theta_\pi})$ in $\bar{\nu}$ Mode}
Coherent and resonant $\pi^0$ production are distinguishable by $\cos{\theta_\pi}$, which is the cosine of the lab angle of the outgoing $\pi^0$ with respect to the beam direction.  It turns out that it is even better to study coherent $\pi^0$'s in terms of the pion energy-weighted angular distribution since in coherent events, $E_\pi(1-\cos{\theta_\pi})$ has a more regular shape as a function of momentum, than $\cos{\theta_\pi}$ alone.  Thus, we will fit for the coherent content as a function of the pion energy-weighted angular distribution.

This fit has in fact been done in neutrino mode [3].  However, in antineutrino scattering there is a helicity suppression for most interactions, including resonant production of $\pi^0$'s, but not for coherent production.  Thus, the ratio of coherent to resonant scattering, which is small, is enhanced in antineutrino running.

\vspace{-0.35cm}
\subsection{Event Selection}
\vspace{-0.1cm}

In selecting MiniBooNE NC $\pi^0$ events, we require the following: only one sub-event (so that the interaction is NC), veto hits<6 (to eliminate cosmic rays), tank hits>200 (to eliminate Michel electrons),  and the event occurs within the fiducial volume. The surviving events are then reconstructed under four hypotheses: a single electron-like Cherenkov ring, a single muon-like ring, two gamma-like rings with unconstrained kinematics, and two gamma-like rings with the invariant mass = $m_{\pi^0}$.  The reconstruction package uses a detailed model of track light production and propagation in the tank to predict the charge and time of hits on each PMT.  Event parameters are then varied to maximize the likelihood of the observed hits.

\vspace{-0.35cm}

\subsection{Data/MC Comparisons}
\vspace{-0.1cm}
The following figures show comparisons between MiniBooNE antineutrino data and MC, which have been relatively normalized.  Only statistical errors are shown.  The MC includes a rescaling of the Rein-Sehgal coherent cross section based on the measurement in [3].  The antineutrino data set shown has not yet been fit for coherent content.  (This is in progress.)\\

\begin{figure}[!h]
\hfill
\begin{minipage}[b]{.45\textwidth}
\begin{center}
  \includegraphics[height=.245\textheight]{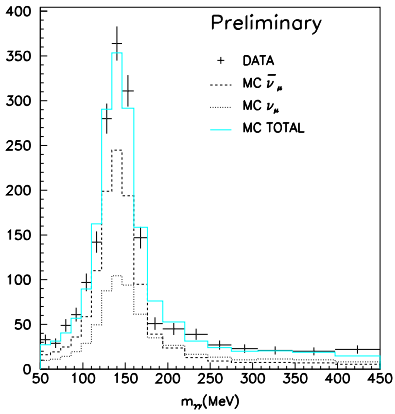}
  \vspace{-.25cm}
  \caption{The invariant mass with the right sign and wrong sign contributions shown.}
\end{center}
\end{minipage}
\hfill
\end{figure}

\begin{figure}[!h]
  \hfill
  \begin{minipage}[b]{.45\textwidth}
    \begin{center}  
      \includegraphics[height=.245\textheight]{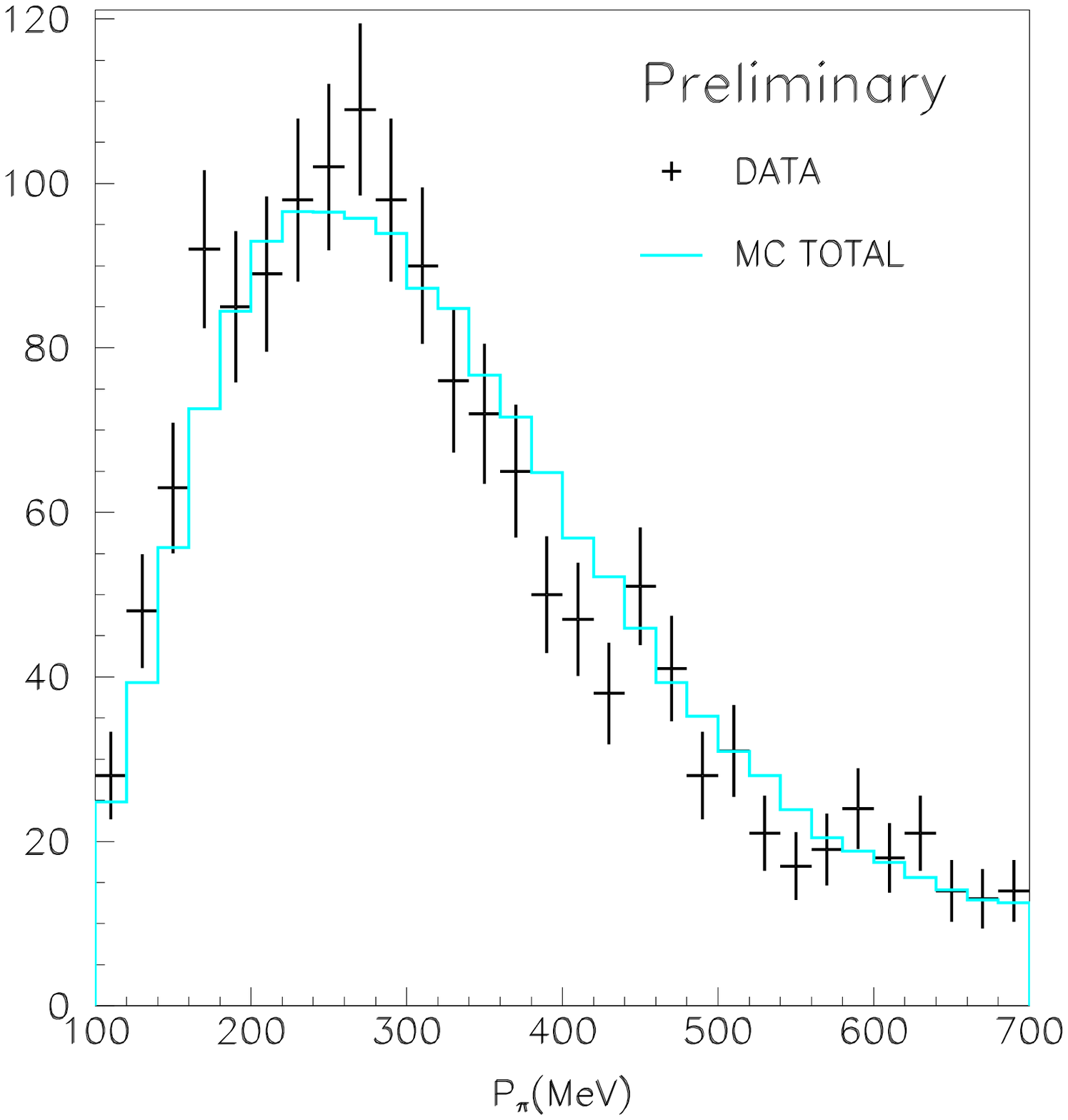}
      \vspace{-.25cm}
      \caption{$\pi^0$ momentum.}
      \label{fig:sub:b}
    \end{center}
  \end{minipage}
  \hfill
  \begin{minipage}[b]{.45\textwidth}
    \begin{center} 
      \includegraphics[height=.245\textheight]{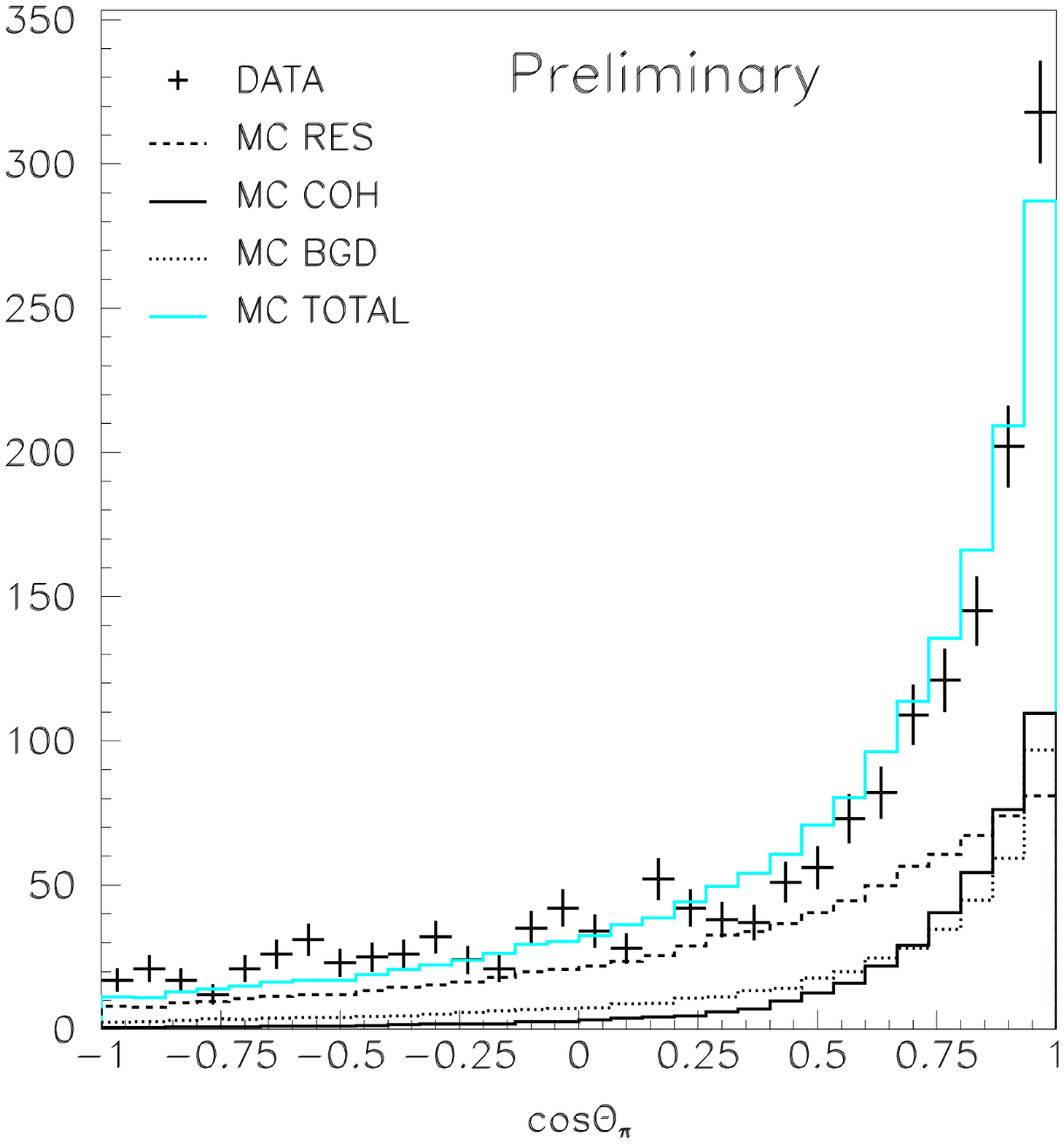}
      \vspace{-.25cm}
      \caption{The left plot shows the $\pi^0$ momentum.  The right plot shows the cosine of the $\pi^0$ angle with respect to the beam direction.  Here, RES denotes $\pi^0$'s produced from a primary interaction delta resonance and COH denotes $\pi^0$'s produced from coherent scattering.  BGD denotes an interaction in which a $\pi^0$ is created via a different means than RES and COH.  MC TOTAL is the sum of RES, COH, and BGD.}
      \label{fig:sub:c}
    \end{center}
  \end{minipage}
  \hfill
\end{figure}

\begin{figure}[!h]
  \hfill
  \begin{minipage}[b]{.45\textwidth}
    \begin{center}  
      \includegraphics[height=.245\textheight]{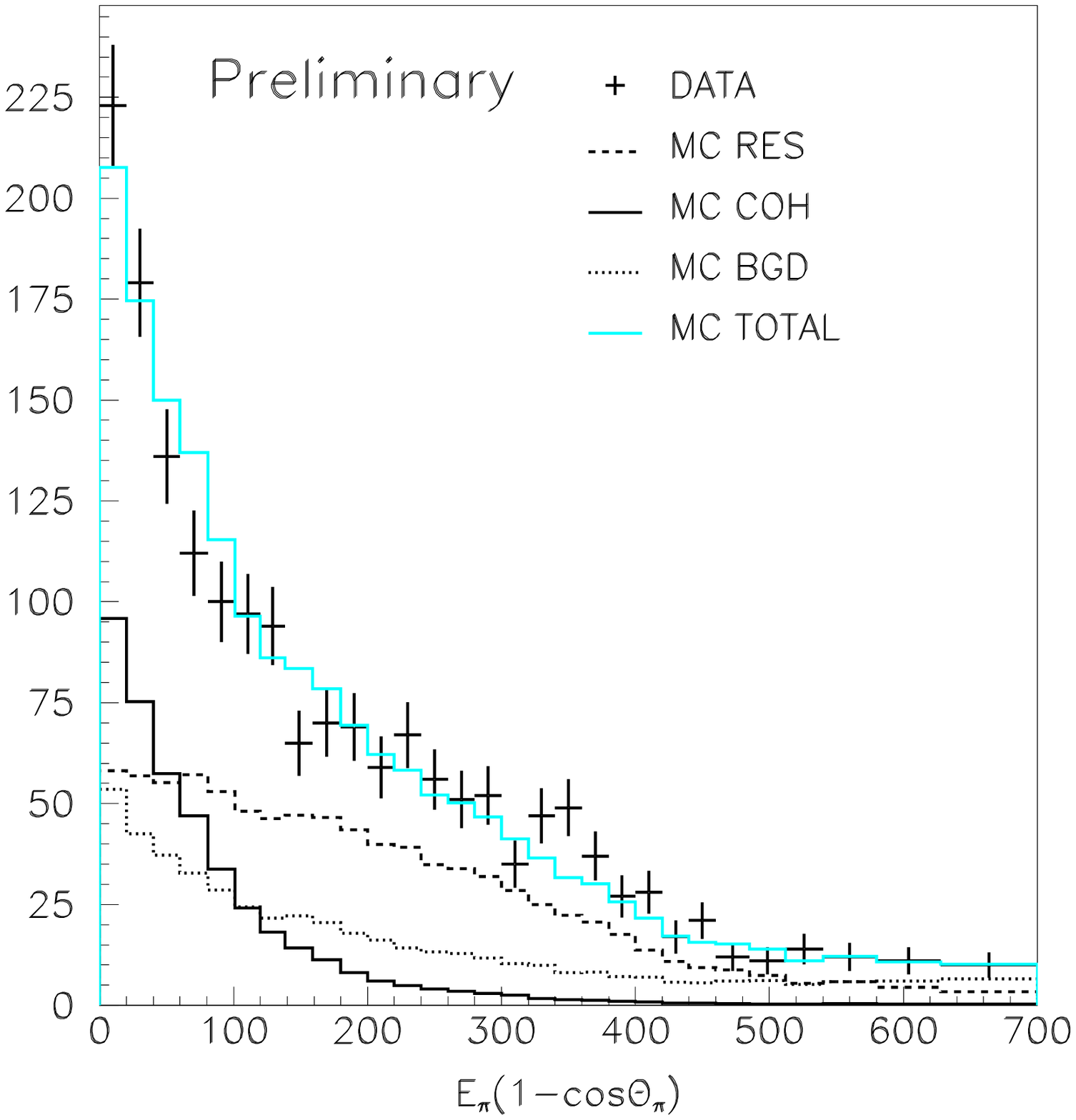}
      \vspace{-.25cm}
      \caption{The pion energy-weighted angular distribution with total MC shown}
      \label{fig:sub:b}
    \end{center}
  \end{minipage}
  \hfill
  \begin{minipage}[b]{.45\textwidth}
    \begin{center} 
      \includegraphics[height=.245\textheight]{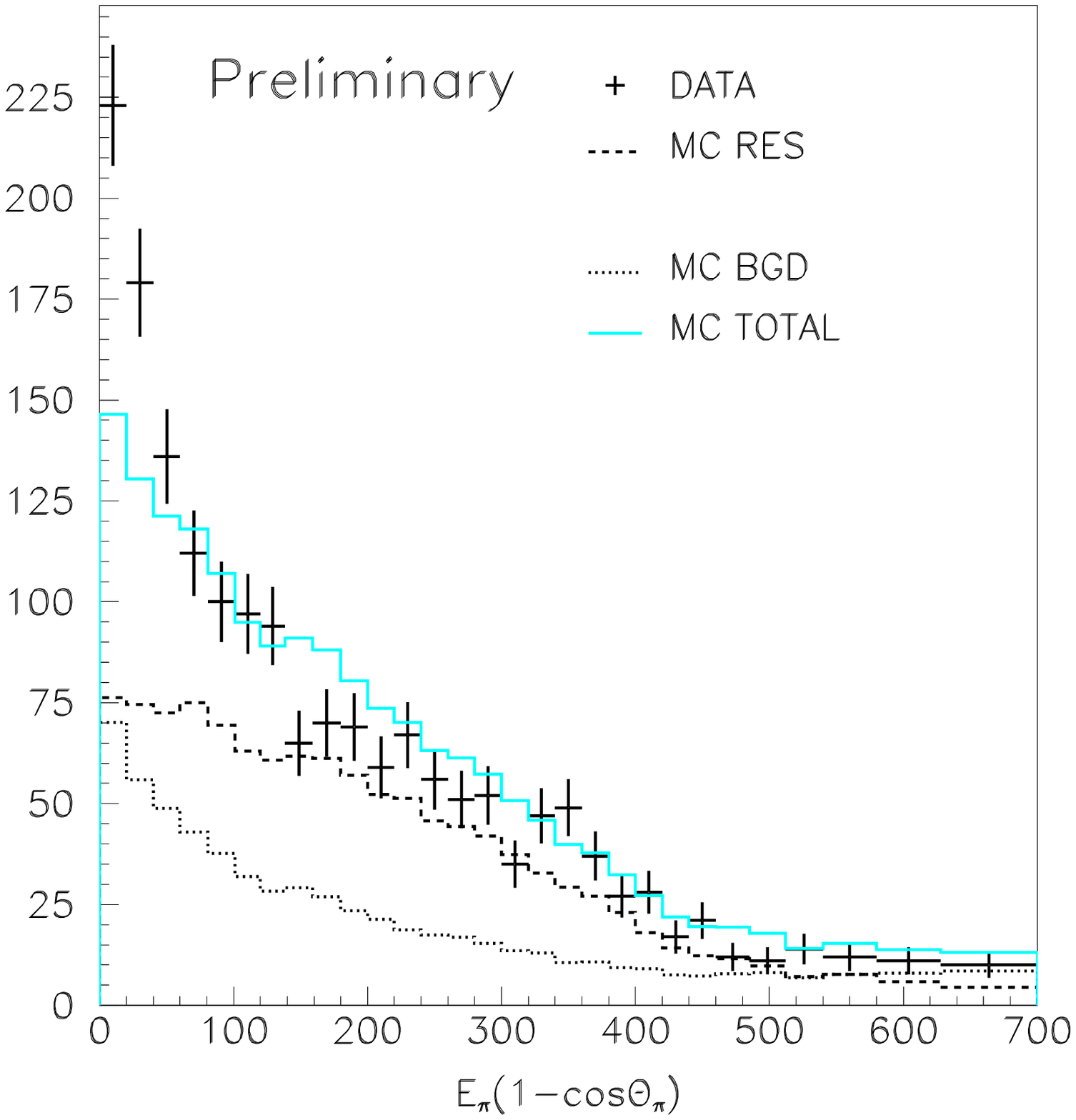}
      \vspace{-.25cm}
      \caption{The left plot shows the pion energy-weighted angular distribution.  The right plot also shows the pion energy-weighted angular distribution but with no coherent contribution in the total MC.  Here, RES, COH, and BGD denote the same quantities described in caption 2.}
      \label{fig:sub:c}
    \end{center}
  \end{minipage}
  \hfill
\end{figure}

When coherent $\pi^0$ production is absent from the MC (right plot of Fig. 3), one can see the poor agreement between data and MC in the lower end of $E_\pi(1-\cos{\theta_\pi})$, where most of the $\pi^0$'s are expected to be produced from coherent scattering.  The agreement becomes good when coherent $\pi^0$ production is included (left plot of Fig. 3).  Thus, MiniBooNE data is already suggesting evidence for antineutrino coherent $\pi^0$ production.

\begin{theacknowledgments}
I would like to give many thanks to the MiniBooNE Collaboration and the NSF.
\end{theacknowledgments}

\end{document}